\documentclass[reqno]{amsart}
\usepackage{latexsym}
\usepackage{amsfonts}
\usepackage{amssymb}
\usepackage{amsmath}
\usepackage{graphicx}
\usepackage{amsthm}
\makeindex

\newcommand{\BR}{\mathbb{R}}
\newcommand{\BC}{\mathbb{C}}

\begin{document}

\title{Spinors and Pre-metric Electromagnetism}

\author{David Delphenich}
\maketitle

\begin{abstract}
The basic concepts of the formulation of Maxwellian 
electromagnetism in the absence of a Minkowski scalar product on spacetime 
are summarized, with particular emphasis on the way that the electromagnetic 
constitutive law on the space of bivectors over spacetime supplants the role 
of the Minkowski scalar product on spacetime itself. The complex geometry of 
the space of bivectors is also summarized, with the intent of showing how an 
isomorphic copy of the Lorentz group appears in that context. The use of 
complex 3-spinors to represent electromagnetic fields is then discussed, as 
well as the expansion of scope that the more general complex projective 
geometry of the space of bivectors suggests.
\end{abstract}

 \footnotetext{AMS Subject Classification 15A66, 15A75, 51P05, 78A02, 78A25}

\section{Introduction}

The role of spinors in electromagnetism has been well established since they 
were first introduced in non-relativistic form by Pauli, and later, in 
relativistic form, by Dirac. In the first case, the necessity came from the 
Stern-Gerlach experiment, which showed that the electron has two distinct spin 
states. In the second case, the necessity came from the purely mathematical 
desire to eliminate the physically puzzling negative energy solutions to the 
Klein-Gordon equation, which was the most logical relativistic formulation 
of the Schr\"{o}dinger equation for quantum wavefunctions. Ironically, the 
resulting relativistic Dirac wave equation did not actually eliminate the 
negative energy solutions, which represented the positron wavefunctions in a 
historical era in which such particles had yet to be discovered.

We denote four-dimensional Minkowski space by $\mathfrak{M}$$_{4}$ = ($\BR$$^{4}$, \textit{$\eta $}), in 
which we have chosen the signature convention \textit{$\eta $ }= diag(+1, --1, 
--1, --1) for specificity. The wavefunctions \textit{$\psi $}: 
M$_{4}$ $\to$ $\BC$$^{4}$ that obey the Dirac equation:
\begin{equation}
(i\gamma ^{\mu } \partial _{\mu }+ m)\psi = 0
\end{equation}

\noindent are then associated with moving charge distributions that define the sources 
of electromagnetic fields.

In this equation, the four Dirac matrices \textit{$\gamma $}$^{\mu }$ , \textit{$\mu $ }= 0, 1, 2, 3 are 
associated with the four corresponding legs of an oriented time-oriented 
orthonormal frame \textbf{e}$_{\mu }$ , \textit{$\mu $ }= 0, 1, 2, 3 on Minkowski space, or 
-- more precisely -- its reciprocal coframe $\theta ^{\mu }$. These 
matrices generate a matrix representation of \textit{Cl}($\BR$$^{4}$, \textit{$\eta $}), the Clifford 
algebra of Minkowski space; the space $\BC$$^{4}$ on which they act is the space 
of field values for the fields on $\mathfrak{M}$$_{4}$ that one calls \textit{Dirac spinors.} Since the group of 
units in \textit{Cl}($\BR$$^{4}$, \textit{$\eta $}) is isomorphic to \textit{SL}(2; $\BC$), in which the proper 
orthochronous Lorentz group can be represented as a subgroup, the Dirac 
spinors carry a representation of the proper orthochronous Lorentz group.

If \textit{$\psi $} represents the wavefunction for a moving electron/positron then one can 
associate it with a conserved current 1-form $J$ whose components relative to 
$\theta ^{\mu }$ are:
\begin{equation}
J_{\mu }=\pm \,ie\bar {\psi }\gamma _\mu \psi ,
\end{equation}

\noindent in which we have lowered the index of by means of the Minkowski scalar 
product.

One always has one -- admittedly trivial -- Clifford algebra that is 
associated with Minkowski space by ignoring its scalar product and giving 
$\BR$$^{4}$ the completely degenerate bilinear form that associates all pairs of 
vectors with zero. The Clifford algebra that is associated with this 
orthogonal structure is simply the exterior algebra of $\BR$$^{4}$, which 
clearly relates to the formulation of Maxwellian electromagnetism.

The introduction the Minkowski scalar product into the definition of the 
Clifford algebra associated with the laws of electromagnetism is intimately 
related to the introduction of the Minkowski scalar product into the 
formulation of the field equations for electromagnetism. Because both 
aforementioned Clifford algebras are defined over the same vector space 
there is an obvious linear isomorphism between the two, namely, the identity 
map; it is not, of course, an algebra isomorphism.

The 2-forms, which 
include the Minkowski field strength 2-form for an electromagnetic field, 
are contained in the even subalgebra of either Clifford algebra. The even 
subalgebra of \textit{Cl}($\BR$$^{4}$; \textit{$\eta $}) is isomorphic to the Clifford algebra 
\textit{Cl}($\BR$$^{3}$, \textit{$\delta $}) of Euclidian $\BR$$^{3}$, which is why the electromagnetic 2-form 
$F$ is often represented in terms of the Pauli matrices \textit{$\sigma $}$_{i}$\c{ } $i $= 1, 2, 3, 
which are usually regarded as non-relativistic objects.

Now, it was first observed by Cartan \cite{Cartan}, and later expanded upon 
by Kottler \cite{Kottler} and Van Dantzig \cite{DVD}, that the only place 
where spacetime metric appears in Maxwell equations is in the Hodge duality 
isomorphisms:
\begin{equation}
\ast : \Lambda ^{\ast }(M) \quad \to \Lambda ^{4-\ast 
}(M), \alpha \mapsto \ast \alpha .
\end{equation}

Kottler and Van Dantzig then succeeded in re-formulating Maxwell's equations 
without the introduction of the usual Lorentzian pseudo-metric, but by 
substituting an electromagnetic constitutive law as the agent of this new 
formulation. (For the sake of brevity, in the sequel, we shall refer to the 
resulting theory of ``pre-metric electromagnetism'' by the acronym PMEM.)

One also must confront the purely physical consideration that the appearance 
and nature of the spacetime pseudo-metric of relativity theory is intimately 
linked with the propagation of electromagnetic waves, even though -- 
paradoxically -- the metric structure of spacetime seems to be ultimately 
more fundamental to the appearance of \textit{gravitational} forces, not electromagnetic ones. 
Hence, there is reason to suspect that the much weaker gravitational force 
is, in some sense, subordinate to the much stronger electromagnetic one. In 
particular, one can define the Lorentzian structure as something that 
appears by way of the principal symbol of the d'Alembertian operator, and 
can be derived from the electromagnetic constitutive law by the use of the 
Fresnel analysis of waves in anisotropic media \cite{Hehl1}, suitably 
adapted to four-dimensional methods.

One gets more fundamental isomorphisms between $\Lambda _{\ast }$($\BR$$^{4})$ 
and $\Lambda $*($\BR$$^{4})$ from Poincar\'{e} duality, which is defined by the 
use of a choice of unit volume element. This suggests that projective 
geometry might be more appropriate for electromagnetism. Another hint in 
this direction comes from the consideration of the symmetries of the 
pre-metric Maxwell equations, in the sense of the symmetries of their space 
of solutions. In work done by the author \cite{DHD1}, it was found that 
although in the absence of deeper analysis there seems to be a choice of 
four possible symmetry groups for PMEM, nevertheless, the one that seems to 
most directly extend the conformal Lorentz symmetry that was established by 
Bateman and Cunningham is the group \textit{SL}(5; $\BR$), which represents the group of 
projective transformations of $\BR$P$^{4}$.

The conformal Lorentz group is associated with the introduction of light 
cones into the tangent spaces of the spacetime manifold. Indeed, physically, 
the measurement of distances in spacetime is facilitated by the introduction 
of electromagnetic waves. However ``most'' bivectors (2-forms, resp.) are 
not wavelike, so the use of a structure -- namely, the Lorentzian structure 
$\textit{g}$ that is associated with the wave solutions restricts the 
generality of Maxwell's equations.

In the next two sections, we shall summarize the basic notions of PMEM, as 
they relate to the issue of spinors and Clifford algebras. We then summarize 
some relevant concepts that pertain to the geometry of $\BR$$^{4}$ when one 
introduces a complex structure on the six-dimensional vector space of 
bivectors (2-forms, resp.). In the central section of the article, we 
address the way that one introduces Clifford algebras into PMEM. Finally, we 
discuss some of the issues that are associated with expanding from Clifford 
algebras to more projective sorts of algebras.

\section{Pre-metric Maxwell equations \cite{Hehl1, DHD1, DHD2}}

We assume that our spacetime manifold $M $is four-dimensional, orientable, and 
given a specific choice \textit{$\varepsilon $} $\in \Lambda ^{4}(M)$ of unit-volume element on 
$T(M)$. One can then define a unit-volume element \textbf{$\epsilon $} $\in $ 
$\Lambda _{4}(M)$ on $T$*($M)$ by choosing the unique 4-vector field 
\textbf{$\epsilon $} such that \textit{$\varepsilon $}(\textbf{$\epsilon $}) = 1. For a natural frame field $\partial _{\mu }=\partial $/$\partial x^{\mu }$ 
that is defined by a local coordinate chart ($U$, $x^{\mu })$ on an open subset 
$U$ in $M$, and its reciprocal local co-frame field \textit{dx}$^{\mu }$, these two volume elements take the local form:
\begin{equation}
\varepsilon = dx ^{0} \wedge dx ^{1} \wedge  dx ^{2} \wedge dx ^{3} = \frac{1}{4!}\varepsilon _{\kappa \lambda \mu \nu }
dx ^{\kappa } \wedge dx ^{\lambda } \wedge  dx ^{\mu } \wedge dx ^{\nu }, 
\end{equation}

\begin{equation}
\epsilon = \partial _{0} \wedge \partial _{1} \wedge \partial _{2} \wedge \partial _{3} = 
\frac{1}{4!} \varepsilon ^{\kappa \lambda \mu \nu } \partial _{\kappa } \wedge 
\partial _{\lambda } \wedge \partial _{\mu } \wedge \partial _{\nu }.
\end{equation}

A first key point of departure from the conventional formulation of 
Maxwell's equations is the fact the divergence operator on $\Lambda _{\ast 
}(M)$, viz., \textit{$\delta $} $\equiv $ {\#}$^{-1}d${\#}, can be defined by the \index{Duality, Poincar\'{e}} \textit{Poincar\'{e} duality} isomorphism:
\begin{equation}
\#: \Lambda _{\ast }(M) \to \Lambda ^{4-\ast }(M), \textbf{a} 
\mapsto  i_{\textbf{a}} \varepsilon = a^{\mu \ldots \nu }
\varepsilon _{\kappa \lambda \mu \nu }.
\end{equation}

\noindent which comes from the volume element, not Hodge duality, which requires a metric. Indeed, this is an important subtlety concerning the divergence 
operator in general, which is often presented as something that requires the 
introduction of a metric. However, one must recall that divergenceless 
vector fields are the infinitesimal generators of local volume-preserving 
diffeomorphisms, which indicates the fundamental relationship between the 
divergence operator and the volume element.

A second point of departure is that the role of an explicitly specified 
electromagnetic constitutive law is given more prominence than in most 
treatments of Maxwell's equations using exterior differential forms.

In general, an \index{electromagnetic constitutive law} \textit{electromagnetic constitutive law} takes the form of an invertible fiber-preserving map:
\begin{equation}
\chi :\Lambda ^{2}(M) \to \Lambda _{2}(M), \quad F \mapsto  \mathfrak{h} = \chi (F),
\end{equation}

\noindent that is a diffeomorphism of the fibers in the nonlinear case and a linear isomorphism:
\begin{equation}
\mathfrak{h}^{\mu \nu }=\tfrac{1}{2}F_{\kappa \lambda }\chi ^{\kappa \lambda \mu \nu }
\end{equation}

\noindent in the linear case. Furthermore, one expects this bundle map to cover the identity, i.e., to take a fiber of $\Lambda ^{2}(M)$ at a given point to a 
fiber of $\Lambda _{2}(M)$ at the same point. The bivector field $\mathfrak{h}$ that corresponds to a given 2-form $F $is referred to as its \index{bivector, electromagnetic excitation}\textit{electromagnetic excitation} bivector field.

If $F$ is the usual \index{two-form, Minkowski} Minkowski 2-form of electromagnetic field strengths, $d$ is 
the exterior derivative operator on $\Lambda $*($M)$, and \textbf{J} is the 
vector field of electric current (i.e., the source of the electromagnetic 
field) then the \index{Maxwell equations, pre-metric} pre-metric Maxwell equations take the form:
\begin{equation}
dF = 0, \quad \delta \mathfrak{h} = \textbf{J }, \quad \mathfrak{h}= \chi (F).
\end{equation}

In local form, these are:
\begin{equation}
\partial _{\lambda}F_{\mu \nu }+ \partial _{\mu}F_{\nu \lambda } 
+ \partial _{\nu}F_{\lambda \mu } = 0, \quad 
\partial _{\mu }\mathfrak{h}^{\mu \nu }=J^{\nu }, \quad
\mathfrak{h} ^{\kappa \lambda } = 
\tfrac{1}{2} \chi ^{\kappa \lambda \mu \nu } F_{\mu \nu }.
\end{equation}

\section{Electromagnetic constitutive laws \cite{Hehl1, DHD2, Post}}

Clearly, the pre-metric Maxwell equations closely resemble the usual Maxwell 
equations, except that the role of the electromagnetic constitutive law has 
supplanted that of the Lorentzian pseudo-metric. Hence, we now briefly 
discuss both physical and mathematical aspects of postulating an 
electromagnetic constitutive law as a fundamental object.

In classical vacuum (Maxwellian) electromagnetism, $\mathfrak{h}$ is linear on the fibers, and if the Minkowski 2-form takes the local form:
\begin{equation}
F =\tfrac{1}{2}F_{\mu \nu } dx^{\mu } \wedge dx^{\nu }=E_{i}
dx^{0} \wedge dx^{i} + \tfrac{1}{2}\varepsilon _{ijk } B^{i} dx^{j} 
\wedge dx^{k}
\end{equation}

\noindent then \textit{$\chi $ }takes the homogeneous isotropic form:
\begin{equation}
\chi (F)=D^{i} \partial _{0} \wedge  \partial _{i} 
+\tfrac{1}{2} \varepsilon ^{ijk } H_{i} \partial _{j} \wedge 
\partial _{k} = 
\varepsilon _{0} \delta ^{ij} E_{j} \partial _{0} \wedge 
\partial _{i} + 
\frac{1}{2\mu _0 }\varepsilon ^{ijk }B_{i} \partial _{j} \wedge
\partial _{k}.
\end{equation}

\noindent in which the indices $i, j, k$ range over the spatial values 1, 2, 3.

The constant \textit{$\varepsilon $}$_{0}$ is referred to as the \index{permittivity, electric} \textit{electric permittivity} (or \index{constant, dielectric} \textit{dielectric constant}) of the vacuum and \textit{$\mu $}$_{0}$ is 
its \index{permeability, magnetic} \textit{magnetic permeability.}

In both of the expressions for $F$ and \textit{$\chi $}($F)$ we have implicitly used the Euclidian 
spatial metric, whose components in the chosen frame are \textit{$\delta $}$_{ij}$, and its 
inverse \textit{$\delta $}$^{ij}$, to raise and lower the index of $B_{i}$ and $H^{i}$, 
respectively. Hence, one should carefully note that the expression for 
\textit{$\chi $}($F)$ is not actually invariant under Lorentz transformations of the local 
frame field $\partial _{\mu }$, but only under Euclidian 
rotation of its spatial members. What makes this intriguing is that the 
speed of propagation for electromagnetic waves in vacuo is derived from 
\textit{$\varepsilon $}$_{0}$ and \textit{$\mu $}$_{0}$:
\begin{equation}
c_{0} = \frac{1}{\sqrt {\varepsilon _0 \mu _0 }}.
\end{equation}

Hence, a fundamental assumption of special relativity -- viz., that 
$c_{0}$ is a constant that is independent of the choice of Lorentz frame -- 
seems to resolve at the pre-metric level to the statement that the constants 
from which $c_{0}$ is constructed can change with the choice of Lorentz 
frame.

One can eliminate the homogeneity restriction and allow \textit{$\varepsilon $} and \textit{$\mu $} to vary with 
position, which is essentially what one does in linear optics. In such a 
case, it is usually not the speed of propagation in the medium that one 
considers, but its index of refraction:
\begin{equation}
n(x)=\frac{c_0 }{c(x)}=\sqrt {\frac{\varepsilon _0 \mu _0 }{\varepsilon 
(x)\mu (x)}}.
\end{equation}

Furthermore, one can drop the isotropy restriction, in which 
case, \textit{$\varepsilon $}$_{0}$ and \textit{$\mu $}$_{0}$ are replaced by 3$\times $3 matrices whose components 
are functions of position. This situation relates to the propagation of 
electromagnetic waves in crystal media, in which the speed of propagation 
can vary with direction, as well as position.

The case of a nonlinear \textit{$\chi $} not only has immediate application to nonlinear 
optics, but also a possible application to effective QED, perhaps in the 
effective models for vacuum polarization, such as the Born-Infeld model. 
However, our immediate concern is with how one introduces Clifford algebras 
into the pre-metric formalism, so we shall concentrate only on the linear 
case.

Returning to the linear case, \textit{$\chi $} also defines a non-degenerate bilinear form 
on $\Lambda ^{2}(M)$:
\begin{equation}
\chi (F, G) \equiv  G(\chi (F)) = \chi _{IJ} F^{I} G^{J} = \tfrac{1}{2} \chi_{\kappa \lambda \mu \nu } F^{\kappa \lambda } G^{\mu \nu }.
\end{equation}

This bilinear form admits a decomposition that is irreducible under the 
action of \textit{GL}(6; $\BR$) on it by congruence (viz., \textit{$\chi $ }$\mapsto $\textit{A$\chi $A}$^{T})$:
\begin{equation}
\chi = ^{(1)}\chi + ^{(2)}\chi +  ^{(3)}\chi ,
\end{equation}

\noindent in which:
\begin{equation}
^{(1)}\chi = \chi - ^{(2)}\chi - ^{3)}\chi
\end{equation}

\noindent is called the \index{constitutive law, electromagnetic, principal part} \textit{principal part}. It is symmetric and ``traceless,'' in the sense that it does 
not contain a contribution that is proportional to the volume element 
\textit{$\varepsilon $}.

The tensor field:
\begin{equation}
^{(2)}\chi = \tfrac{1}{2}(\chi - \chi ^{T})
\end{equation}

\noindent is the anti-symmetric\index{constitutive law, electromagnetic, skewon part} \textit{skewon} part of \textit{$\chi $}. It is associated with established physical phenomena, such as the Faraday effect, and natural optical activity \cite{Post, LLP, Hehl2, Hehl3}, so it is not a purely abstract generalization to include it in \textit{$\chi $}.

The tensor field:
\begin{equation}
^{(3)}\chi = \chi _{0} (E_{I}, E_{I}) \varepsilon
\end{equation}

\noindent is the \index{constitutive law, electromagnetic, axion part} \textit{axion} part of \textit{$\chi $}, which proportional to volume element. It does not affect 
the propagation of electromagnetic waves, but Lindell and Sivola 
\cite{Lindell} have suggested that it might still play a role in some 
electromagnetic media.

In the language of projective geometry, the case of a general \textit{$\chi $} defines a 
\textit{correlation} on the fibers of $\Lambda ^{2}(M)$ (more precisely, their 
projectivizations), namely, a linear isomorphism from each fiber of $\Lambda 
^{2}(M)$ to its dual vector space, which is a fiber of $\Lambda 
_{2}(M)$. A symmetric correlation is called a \textit{polarity} and defines a quadratic 
form. An anti-symmetric correlation defines a \textit{null polarity}, much like a symplectic form 
on an even-dimensional vector space. One can always polarize a correlation 
into a symmetric and an anti-symmetric part, although the individual parts 
do not have to both be non-degenerate.

The manner by which \textit{$\chi $} gives rise to a Lorentzian metric on $T(M)$ follows from 
adding certain restricting assumptions on \textit{$\chi $}. Essentially, one looks for an 
``exterior square root'' \textit{$\chi $} = ``$g \wedge g$,'' since, in the case of the Hodge * 
isomorphism, the role of \textit{$\chi $} is played by the map whose tensor components are:
\begin{equation}
\tfrac{1}{2}(g_{\kappa \lambda } g_{\mu \nu } - g_{\kappa \nu } 
g_{\mu \lambda })
\end{equation}

Physically, the absence of \index{birefringence} birefringence is often a necessary and sufficient 
restricting assumption for the reduction to take place. Birefringence is an 
optical phenomenon that takes the form of the speed of light in a medium -- 
hence, the index and angle of refraction -- depending on the polarization 
direction of the light wave \cite{LLP}.

\section{Geometry of bivectors \cite{DHD3}}

One might say that the basic theme of PMEM is that one must make a shift of 
emphasis from considering the geometry of $M$ by way of a metric $g $on $T(M)$ to 
considering the geometry $M $by way of the various structures that one defines 
on $\Lambda ^{2}(M)$. Hence, we now summarize some of those geometric 
structures. Furthermore, we restrict our scope to the manifold R$^{4}$, 
which amounts to considering a single tangent space to a more nonlinear 
manifold.

The volume element \textit{$\varepsilon $ }defines a real scalar product of signature type (3, 3) on 
$\Lambda _{2}$($\BR$$^{4})$:
\begin{equation}
<\textbf{F}, \textbf{G}> \equiv \varepsilon (\textbf{F} \wedge \textbf{G}).
\end{equation}

The \index{bivectors, isotropic} \textit{isotropic} bivectors of this scalar product -- i.e., $<$\textbf{F}, \textbf{F}$>$ 
= 0 -- define the \index{Klein hypersurface} \textit{Klein hypersurface }in $\Lambda _{2}$($\BR$$^{4})$. A bivector 
\textbf{F} is isotropic iff it is \index{bivectors, decomposable} \textit{decomposable}, where \textbf{F} decomposable iff 
\textbf{F} = \textbf{a} \^{} \textbf{b} for some \textbf{a}, 
\textbf{b}$ \in \BR ^{4}$.

When we express \textbf{F} as $E^{i }$\textbf{E}$_{i}+B^{i 
}$*\textbf{E}$_{i}$ we find that:
\begin{equation}
<\textbf{F}, \textbf{F}> = \textbf{E}^{2} - \textbf{B}^{2}.
\end{equation}

We define the isomorphism \textit{$\kappa $}: $\Lambda _{2}$($\BR$$^{4}$) $\to 
\Lambda _{2}$($\BR$$^{4})$, \textit{$\kappa $} = {\#}\textit{$\chi $}, which we 
\textit{assume} to be proportional to a \index{bivectors, complex structure on space of} complex structure * on $\Lambda _{2}$($\BR$$^{4})$:
\begin{equation}
\kappa ^{2} = - \lambda ^{2} I, \ast \equiv  \lambda ^{-1}\kappa .
\end{equation}

Hence, by definition, *$^{2}= - I$, which is also a property of 
the Hodge * when it acts on 2-forms on a four-dimensional Lorentz manifold. 
However, in the present case, we have not defined the isomorphism $\Lambda 
_{1}$($\BR$$^{4})\mathfrak{g}\to \Lambda _{3}$($\BR$$^{4})$ that Hodge 
duality defines. Interestingly, the remaining isomorphisms $\Lambda 
_{0}$($\BR$$^{4})\to \Lambda _{4}$($\BR$$^{4})$ and $\Lambda 
_{0}$($\BR$$^{4})\to \Lambda _{4}$($\BR$$^{4})$ can still be 
defined in the absence of a metric. One assigns the real number $a$ to the 
4-vector $a$\textbf{$\epsilon $}, and the 4-vector$ a$\textbf{$\epsilon $} to 
the real number -- $a$, respectively. Hence, we have defined Hodge duality only 
on the even subalgebra of $\Lambda _{\ast }$($\BR$$^{4})$, namely:

\begin{equation}
\Lambda _{+}(\BR ^{4}) = \Lambda _{0}(\BR ^{4}) 
\oplus  \Lambda _{2}(\BR ^{4}) 
\oplus  \Lambda _{4}(\BR ^{4}) 
= \BR \oplus \Lambda _{2}(\BR ^{4}) \oplus \BR \epsilon .
\end{equation}

In order to complete the isomorphism of $\Lambda _{2}$($\BR ^{4}$) with 
$\BC ^{3}$, we must give it a complex scalar multiplication, which we define 
by:
\begin{equation}
(\alpha + i \beta )\textbf{F} \equiv  \alpha \textbf{F} + \beta \ast \textbf{F }
\end{equation}

A (non-canonical) $\BC$-linear isomorphism from $\Lambda _{2}$($\BR$$^{4})$ to 
$\BC$$^{3}$ is then defined by a choice of complex 3-frame 
on $\Lambda _{2}$($\BR ^{4}$). That is, if:
\begin{equation}
\textbf{F} = E^{i} \textbf{E} _{i} + B^{i} \ast \textbf{E} _{i} = (E^{i} + 
{iB}^{i}) \textbf{E}_{i},
\end{equation}

\noindent as above, and the complex 3-frame \textbf{E}$_{i}$ on $\Lambda 
_{2}$($\BR ^{4}$) corresponds to the canonical 3-frame on $\BC ^{3}$ then the 
vector in $\BC ^{3}$ that corresponds to \textbf{F} has the complex components 
$E^{i}$ + \textit{iB}$^{i }$.

Since any choice of 3-frame, such as \textbf{E}$_{i }$, defines a direct sum 
decomposition of $\Lambda _{2}$($\BR ^{4})$ into essentially a ``real'' 
subspace and an ``imaginary'' subspace, this decomposition is clearly not 
unique; in fact, it closely analogous to a 3+1 decomposition of $\BR ^{4}$, 
although we refer the curious to a more detailed treatment \cite{DHD3} by 
the author.

Note further that we can extend the $\BC$-linear isomorphism of $\Lambda 
_{2}$($\BR ^{4})$ with $\BC ^{3}$ to a $\BC$-linear isomorphism of the even 
subalgebra of $\Lambda _{2}$($\BR ^{4})$ with $\BC ^{4}$, by simply regarding 
the $\BR$ $\oplus $ $\BR$\textbf{$\epsilon $} part as corresponding to $\BC$ by the isomorphism $a $+ *$b \quad \mapsto  \quad a$ + \textit{ib}.

Since we now regard * as the fundamental object on $\Lambda 
_{2}$($\BR$$^{4})$, we should consider the subgroup of \textit{GL}(6; $\BR$) that preserves 
*. One finds that this subgroup is isomorphic to \index{group, complex general linear} \textit{GL}(3; $\BC$). Its elements can be expressed as either 3$\times $3 complex invertible matrices or as 3+3-partitioned 6$\times $6 real invertible matrices by the association:

\begin{equation}
A + iB \leftrightarrow  \quad 
\left[ {{\begin{array}{*{20}c}
 A \hfill & {-B} \hfill \\
 B \hfill & A \hfill \\
\end{array} }} \right]=\left[ {{\begin{array}{*{20}c}
 A \hfill & 0 \hfill \\
 0 \hfill & A \hfill \\
\end{array} }} \right]+\ast \left[ {{\begin{array}{*{20}c}
 B \hfill & 0 \hfill \\
 0 \hfill & B \hfill \\
\end{array} }} \right],
\end{equation}

\noindent in which we are representing the matrix of * by:

\begin{equation}
\ast = \left[ {{\begin{array}{*{20}c}
 0 \hfill & {-I} \hfill \\
 I \hfill & 0 \hfill \\
\end{array} }} \right].
\end{equation}

Although, as noted above, there is good reason to keep the isomorphism 
\textit{$\chi $} more general, nevertheless, \textit{when $\chi $ is symmetric}, the *-isomorphism defines another real scalar product on $\Lambda _{2}$($\BR$$^{4})$:
\begin{equation}
(\textbf{F}, \textbf{G}) = <\textbf{F}, *\textbf{G}> .
\end{equation}

Relative to the frame that we have been habitually using we have:
\begin{equation}
(\textbf{F}, \textbf{F}) = 2\textbf{E} \cdot \textbf{B}.
\end{equation}

Combining both of the aforementioned real scalar products defines a \index{bivectors, complex orthogonal structure on space of} 
Euclidian structure on $\Lambda _{2}$($\BR$$^{4})$ by way of
\begin{equation}
<\textbf{F}, \textbf{G}>_{\BC } = (\textbf{F}, \textbf{G}) + i <\textbf{F}, 
\textbf{G}> .
\end{equation}

The subgroup of \textit{GL}(3; $\BC$) that preserves this scalar product is isomorphic to 
$O$(3; $\BC$). The introduction of a unit-volume element on $\Lambda 
_{2}$($\BR$$^{4})$, which is straightforward, then defines a reduction to 
\textit{SO}(3; $\BC$), which is isomorphic to \textit{SO}$_{0}$(3, 1). This isomorphism really 
embodies the essence of the reduction from the geometry of $\Lambda 
_{2}$($\BR$$^{4})$ given the * isomorphism, which is essentially the geometry 
of $\BC$$^{3}$, to the geometry of $\BR$$^{4}$ with the Lorentz scalar product.

\section{Bivector Clifford algebras}

In order to define a \index{bivectors, Clifford algebras on space of} Clifford algebra -- whether real or complex -- one must 
have a vector space with an orthogonal structure -- i.e., a scalar product 
$\mathfrak{t}$ defined on it. The vector space in question at the moment is 
$\Lambda _{2}$($\BR$$^{4})$, which we have just observed can be regarded as 
having six real dimensions or three complex dimensions; we have also defined 
a scalar product in either case.

When $\Lambda _{2}$($\BR$$^{4})$ is regarded as a real vector space, one can 
use the real scalar product $<$.,.$>$, whose orthogonal group has \textit{SO}(3, 3) as 
its identity component. This 64-dimensional real Clifford algebra has been 
dealt with by Harnett \cite{Harnett}. It is representable by the matrix 
algebra End($\BR ^{4} \oplus \BR ^{4}$*). These matrices generalize the Dirac 
matrices that one uses to represent the Clifford algebra of Minkowski space. 
However, when one gives $\Lambda _{2}$($\BR ^{4})$ a complex structure, the 
role of \textit{SO}(3, 3) does not seem as physically meaningful as that of \textit{SO}(3, $\BC$). 
Hence, we shall not elaborate on this example at the moment.

By contrast, when $\Lambda _{2}$($\BR ^{4})$ is regarded as a complex vector 
space, we can use the complex orthogonal structure $<$.,.$>_{\BC}$, whose 
orthogonal group has the identity component \textit{SO}(3, $\BC$). The resulting complex 
eight- (real sixteen-) dimensional Clifford algebra is, by definition, 
\textit{Cl}($\BR$$^{3}$, \textit{$\delta $}), which is $\BR$-isomorphic to \textit{Cl}($\BR$$^{4}$, \textit{$\eta $}) -- i.e., the 
Clifford algebra of Minkowski space -- and $\BC$-isomorphic to the 
matrix algebra M$_{2}$($\BC$) $\oplus $ M$_{2}$($\BC$). Consequently, these matrices are representable by the usual Dirac matrices.

One notes that a key difference between using the complex Clifford algebra 
\textit{CL}($\BC ^{3}$, \textit{$\delta $}) and the $\BR$-isomorphic real Clifford algebra \textit{Cl}($\BR ^{4}$, \textit{$\eta $}) is that the most natural spinor fields that are associated with the real 
form take their values in $\BC ^{4}$, but the $\BC$-isomorphism of $\Lambda 
_{2}$($\BR ^{4})$ with $\BC ^{3}$ suggests that the most natural spinor fields, 
at least from the standpoint of the electromagnetic field strength 2-form, 
are fields that take their values in $\BC ^{3}$; these fields are sometimes 
referred to as \index{spinors, three-component} ``three-component spinors \cite{Peres, CDD},'' and seem to be 
fundamental in the complex formulation of general relativity. Like the Dirac 
spinors, they still carry a representation of the proper orthochronous 
Lorentz group, by way of its isomorphism with either an \textit{SO}(3; $\BC$) subgroup of 
\textit{GL}(3; $\BC$) or an \textit{SL}(2; $\BC$) subgroup.

Although, as just observed, the representation of electromagnetic field 
2-forms by 3-spinor fields is immediate when one chooses a complex 3-frame 
for $\Lambda _{2}$($\BR ^{4})$, nevertheless, the use of 4-spinors is more 
related to the representation of the wave function for the source current 
\textbf{J}. In particular, the two representations of the Lorentz group have 
different weights (i.e., spins). Hence, one must still determine the manner 
by which one represents \textbf{J} in terms of things derivable from complex 
3-spinors.

Since Dirac 4-spinors are really Cartesian products of 2-spinors, 
corresponding to the isomorphism of \textit{Cl}($\BR ^{4}$, \textit{$\eta $}) with M$_{2}$($\BC$) $\oplus $ M$_{2}$($\BC$), one might investigate the representation of 4-spinors as 
tensor products of 2-spinors; although the space of such tensor products is 
also four-complex-dimensional, nevertheless, the physical interpretation of 
such a construction would be that they are bound states of more elementary things. Furthermore, as 
we saw above, we have a $\BC$-linear isomorphism of $\BC ^{4}$ with the even 
subalgebra of $\Lambda _{\ast }$($\BR ^{4})$. However, this suggests a 
natural decomposition of $\BC ^{4}$ into $\BC \oplus \BC ^{3}$, whereas the more 
established decompositions of Dirac spinors are into pairs of 2-spinors, 
which suggests $\BC ^{2} \oplus \BC ^{2}$.

Finally, along the same lines, one notes that if one defines the complex 
conjugation operator on $\Lambda _{2}$($\BR ^{4})$ that corresponds to the 
complex conjugation operator on $\BC ^{3}$ by way of the chosen isomorphism, 
then the complex orthogonal structure also defines a \index{bivectors, Hermitian structure on space of} Hermitian structure, 
i.e., a Hermitian inner product, by way of:
\begin{equation}
({\bf F}, {\bf G}) =<{\rm {\bf F}},{\rm {\bf \bar {G}}}>_\BR .
\end{equation}

Consequently, one can reduce the \textit{SO}(3; $\BC$) subgroup of \textit{SL}(3; $\BC$) to \textit{SU}(3). Although 
this group represents the color gauge symmetry of the strong interaction and 
the components of a complex 3-spinor field can represent the $u, d, $and $s$ quarks, clearly, the details of how this mathematical coincidence relates to 
physical theory needs to be developed further. One hint is that the 
quadratic form that is associated with the Hermitian structure takes the 1+3 
form:
\begin{equation}
(\textbf{F}, \textbf{F}) = \textbf{E} ^{2} + \textbf{B} ^{2},
\end{equation}

\noindent which is proportional to the electromagnetic field Hamiltonian. Hence, unitary transformations of $\BC$$^{3}$, given the Hermitian structure just 
defined, would preserve the field Hamiltonian. Whether this leads to a more 
direct route to the unification of the theories of the electromagnetic and 
strong interactions is a worthy point to ponder.

\section{Role of projective geometry in PMEM}

So far, we have considered how the principal part of the linear 
electromagnetic constitutive tensor \textit{$\chi $ }defines an orthogonal structure on the 
space of bivectors over $\BR$$^{4}$. In light of the fact that apparently the 
existence of a non-vanishing skewon contribution to the electromagnetic 
constitutive law is not a physical triviality, apparently a complete 
analysis of pre-metric electromagnetism must confront the role of that 
constitutive law as a correlation, not merely a metric. Hence, one 
conjectures that the most appropriate geometrical context for pre-metric 
electromagnetism might be projective geometry -- indeed, 
\index{geometry, complex projective} \textit{complex} projective geometry.

One suggestive result in this direction was derived by the author in an 
analysis of the symmetries of pre-metric Maxwell system \cite{DHD1}. Just 
as the Maxwell equations in metric form admitted the conformal Lorentz group 
as a symmetry group that acts on their space of solutions, the pre-metric 
Maxwell equations seem to give a more ambiguous result: that the symmetry 
group is either the affine group for $\BR ^{4}$, the group of projective 
transformations of $\BR$P$^{4}$, the group of diffeomorphisms of $\BR ^{4}$ whose volume varies as \textit{t}$^{3}$, or the full diffeomorphism group. One's intuition is that it is the group of projective transformations that is the most physically meaningful expansion from the conformal Lorentz group that Bateman and Cunningham established.  However, the example of the expansion of a spherical wavefront shows that the transformations whose infinitesimal generators have constant divergence might also have a certain physical appeal.

As for the suggestion that it is complex projective geometry that we are 
concerned with, note that a \index{duality plane} \textit{duality plane} in $\Lambda _{2}$($\BR ^{4})$ -- 
viz., a space spanned by some bivector \textbf{F} and *\textbf{F} -- 
corresponds with a complex line in $\BC ^{3}$ under a $\BC$-isomorphism of $\Lambda 
_{2}$($\BR ^{4})$ with $\BC ^{3}$. Hence, the space of duality planes in 
$\Lambda _{2}$($\BR ^{4})$ is projectively equivalent to $\BC$P$^{2}$. 
Physically, a duality plane in $\Lambda _{2}$($\BR ^{4})$ is also related to 
the polarization plane in $\BR ^{4}$ that is defined by \textbf{F} when it is 
isotropic.

The matter of what ``geometric algebra'' might represent projective geometry 
is possibly resolved when one remembers that Grassmann's intent in defining 
what is now called the exterior -- or Grassmann -- algebra of a vector space 
was precisely that of representing the incidence relations between linear subspaces 
in terms of algebraic operators on elements of a vector space. Of course, 
the more physically fundamental question is: what happens when one restricts 
oneself to projective transformations that preserve a given correlation, 
i.e., linear electromagnetic constitutive law \textit{$\chi $}? This suggests an expansion 
of the Lorentz group to a more complicated linear algebraic group whose 
algebraic structure would depend upon the properties of \textit{$\chi $}.

This expansion of the scope of the spinors in electromagnetism from Dirac 
spinors to projective spinors begs some further questions:

Insofar as conventional spinors are \textit{wave}functions, what do spinor fields 
represent in the absence of the basis for defining waves, viz., light cones? 
Of course, the moving charge distributions that generate electromagnetic 
fields are always massive, so one might still be able to define the 
wavefunctions of massive matter, even if one could not define photons.

As pointed out in the introduction, Dirac spinors carry a representation of 
the proper orthochronous Lorentz group \textit{SO}$_{0}$(1,3), by way of its universal 
covering group \textit{SL}(2; $\BC$). Which group might one expect projective spinors to 
carry a representation of: \textit{SL}(5; $\BR$), which defines the projective 
transformations of $\BR$P$^{4}$, or \textit{SL}(3; $\BC$), which defines the projective 
transformations of $\BC$P$^{2}$? From the preceding discussions, one suspects 
that it is the complex projective group that plays the fundamental role, but 
one should recall that the essence of the projective relativity 
\cite{Schmutzer} that grew out of the Kaluza-Klein program was the embedding of 
the spacetime manifold in $\BR ^{5}$ in a projective manner. Hence, the real 
projective group might also be of interest.

\textbf{Acknowledgements} The author wishes to thank Gerald Kaiser for the 
invitation to present a paper at this conference, Pierre Angles and the 
SUPAERO facility for their hospitality and able conference administration, 
and Bethany College for providing a congenial and unhurried research 
environment.

\small
\vskip 1pc
{\obeylines
\noindent David Delphenich
\noindent Physics Department
\noindent Bethany College
\noindent Lindsborg, KS 67456 USA
\noindent E-mail: {delphenichd@bethanylb.edu}
}
\noindent Submitted: October 27, 2005, Revised: TBA

\clearpage
\pagestyle{empty}
\hspace{2.in}
\end{document}